\documentclass[amsmath,amssymb,aps,prd, superscriptaddress, preprintnumbers,nofootinbib,a4paper, 11pt]{revtex4-1}
\usepackage{amsthm}
\usepackage{color}
\usepackage{dcolumn}
\usepackage{bm}
\usepackage{bbm}
\usepackage{pxfonts}
\usepackage{slashed} 
\usepackage{url}
\usepackage{graphicx,rotating,colortbl}

\usepackage[ margin=5pt, font=normalsize,labelfont=bf,justification=raggedright]{caption}

\usepackage[bookmarks, pagebackref=false]{hyperref}
\definecolor{rossoCP3}{cmyk}{0,.88,.77,.40}
\definecolor{verdeCP3}{rgb}{0.09765625, 0.57421875, 0.1015625}
\definecolor{bluCP3}{rgb}{0, 0.23, 0.67}

\newcommand{\GM}{0.06}

\newcommand{\Excess}{750~{\rm GeV}}

\newcommand{\fb}{\,{\rm fb}}

\newcommand{\ea}[1]{
\begin{align}
#1
\end{align}
}

\newcommand{\be}{\begin{eqnarray}}
\newcommand{\ee}{\end{eqnarray}}

\newcommand{\change}[1]{{  #1}}

\begin{document}
\title{\Large  \color{rossoCP3} ~~ \\
A natural Coleman-Weinberg  theory explains the diphoton  excess }
\author{Oleg Antipin}
\email{oantipin@irb.hr} 
\affiliation{Rudjer Boskovic Institute, Division of Theoretical Physics, \\Bijeni\v cka 54, HR-10000 Zagreb, Croatia}
\author{Matin Mojaza}
\email{mojaza@nordita.org} 
\affiliation{NORDITA, KTH Royal Institute of Technology and Stockholm University, Roslagstullsbacken 23, SE-10691 Stockholm, Sweden}
\author{Francesco  Sannino}
\email{sannino@cp3.dias.sdu.dk} 
\affiliation{ CP$^{3}$-Origins \& Danish IAS, University of Southern Denmark, Campusvej 55, DK-5230 Odense M, Denmark}

\begin{abstract}
It is possible to delay the hierarchy problem, by replacing the standard Higgs-sector by the Coleman-Weinberg mechanism, and at the same time ensure perturbative naturalness through the so-called Veltman conditions. As we showed in a previous study, minimal models of this type require the introduction of an extra singlet scalar further coupled to new fermions. In this constrained setup the Higgs mass was close to the observed value and the new scalar mass was below a TeV scale. 
Here we first extend the previous analysis by taking into account the important difference between running mass and pole mass of the scalar states. We then investigate whether these theories can account for the 750 GeV excess in diphotons observed by the LHC collaborations.  New QCD-colored fermions in the TeV mass range coupled to the new scalar state  are needed to describe the excess. We further show, by explicit computation of the running of the couplings, that the model is under perturbative control till just above the masses of the heaviest states of the theory.  We further suggest related testable signatures and thereby show that the LHC experiments can test these models. 
\\[.1cm]
{\footnotesize  \it Preprint: CP$^3$-Origins-2015-56 DNRF90, DIAS-2015-56}
 \end{abstract}

\maketitle

\thispagestyle{empty}
\newpage

\section{ Introduction }
It is possible that the tension between the experimental value of the Higgs boson mass and the naturalness principle is alleviated by cancellation of the additive corrections to the mass at the perturbative level.
It is well-known that this happens to all-orders in perturbation theory by supersymmetry. 
However, it may also be that cancellations happen only to a finite order, and thus are of accidental nature.
In such a case, the hierarchy problem is not solved, but delayed to higher scales.
This paradigm was first investigated by Veltman in Ref.~\cite{Veltman:1980mj}, long before the discovery of the top-quark and Higgs-boson, where by eliminating the one-loop additive corrections to the Higgs mass, predictions for the masses of the top-quark and Higgs-boson were derived, which we today know are not compatible with data. However, extensions of the Standard Model (SM) introduce new degrees of freedom that may lead to compatibility between delayed naturalness and experimental data.
In Ref.~\cite{Antipin:2013exa} we pursued this paradigm within the context of electroweak symmetry breaking through the Coleman-Weinberg mechanism, where we showed that the paradigm in many, but one, examples is not compatible with experimental data. 
The only viable example turns out to be a simple one, where the Higgs-sector is extended only by a real singlet scalar field, \change{which also has couplings to a new fermionic sector}.

In this paper, we revisit this model, where in Sec.~\ref{review} we go through the details of the Coleman-Weinberg mechanism. 
In Sec.~\ref{polemass}, we add details on the physical predictions of masses and couplings of the model. In Sec.~\ref{predictions}, we specify the new fermionic sector, and study the phenomenological signatures, explicitly showing that a diphoton excess at 750 GeV \cite{Atlas, CMS:2015dxe} is \change{consistent within the Coleman-Weinberg paradigm and leads in this model to new predictions for the mass and couplings of the new fermions.
In Sec.~\ref{predictions} we discuss other predictions that are consequences of the model that could be tested by the experiments.
Finally, in Sec.~\ref{conclusions} we discuss the Veltman conditions for delayed naturalness, comment on the UV behavior of the model by studying the running couplings, and give our final conclusions.}

\section{Electroweak Symmetry Breaking through a Perturbatively Natural Coleman-Weinberg Mechanism}
\label{review}

Following the paradigm explained in the introduction above, we discovered in Ref.~\cite{Antipin:2013exa} an intriguing extension and modification of the SM where, at the one-loop level, one predicts a Higgs-boson with the observed 126 GeV value of the mass, while predicting yet another massive scalar with running mass value of about 540 GeV, and a coupling to a new fermionic sector, through the condition of delayed naturalness. In this work we revisit the model from the phenomenological perspective, and thus instead use all known experimental data as input, and study the consequences on the model and for delayed naturalness. We start by summarizing the Coleman-Weinberg scenario, which is here replacing the usual Higgs-mechanism.

By replacing the standard Higgs-potential with an approximately flat potential, it was shown long time ago by Coleman and Weinberg~\cite{Coleman:1973jx} that quantum corrections induce spontaneous symmetry breaking, leading to a prediction for the Higgs-boson mass of around $5$ GeV.
The flatness of the potential requires a vanishing of the mass-term before symmetry breaking and nearly vanishing of the Higgs-self-coupling at a specific scale. This is the constraint that returns a prediction of the mass after symmetry breaking.
Specifically, the interacting Lagrangian of the SM Higgs-sector reads in this scenario:
\ea{
\mathcal{L}_{I,\mu_0}^{H} = \lambda \left (H^\dagger H \right)^2  - \frac{1}{2}\left (g^2 W_\mu^+ W^{ - \mu} + \frac{g^2+{g^\prime}^2}{2} Z_\mu Z^\mu \right ) H^\dagger H + y_t (\bar{t}_L , 0) \left(i\sigma^2 H^*\right) t_R + {\rm h.c.} +  \
\text{c.t.} \ ,
}
where the renormalized mass is set to zero and we are neglecting the couplings to the leptons and light quarks. The parameters are considered to be renormalized and we have thus added corresponding counterterms (c.t.). The Coleman-Weinberg (CW) mechanism works, if $\lambda =\lambda(\mu_0)$ is small compared to the other couplings at the scale $\mu_0$, i.e.
\ea{
\lambda(\mu_0) \approx 0 + \mathcal{O}(g^4, {g^\prime}^4, y_t^2) \ ,
}
in which case the quantum (one-loop) corrections to the Higgs-potential become relevant and generically induce a non-zero vacuum expectation value for the Higgs field.

On the other hand, one-loop corrections also induce corrections to the renormalized Higgs-mass, proportional to the cutoff in a cutoff regularized computation of the Higgs-potential, or to the highest mass scale in a dimensionally regularized computation, leading in any case to a hierarchy problem.
One may readjust the counterterms to balance out these corrections, however, this leads to the issue of fine-tuning in the sense that readjustment of counterterms have to be done every time higher-order corrections are computed. One way to avoid this is through the Veltman conditions. If the cutoff or the new mass scale is not too high and the one-loop corrections to the mass cancel each other, then no fine-tuning is needed. This is the paradigm we would like to establish. 
In Ref.~\cite{Antipin:2013exa} we examined several scalar extensions of the SM within this paradigm, and showed that most, but one, are ruled out by experiment. 
The working extension consists of one real scalar singlet $S$, which also has to couple to some fermionic sector. The additional interaction in the Lagrangian reads:
 \be
\mathcal{L}_{I,\mu_0}^S= 
\lambda_{HS} H^\dagger H \  S^2+ \frac{\lambda_S}{4} S^4
+ Y_{ij} S \psi_i \psi_j + \text{h.c.} + \text{c.t} \ ,
\label{lagr}
\ee
where $Y_{ij}$ denotes an as yet unspecified fermionic sector.
The constraint from requiring the potential to be bounded from below is found by completing the square and reads:
\ea{
\label{bounded}\lambda \geq 0 \ , \quad \lambda_S \geq 0 \ ,  \quad  \text{and if \ $\lambda_{HS}<0$ :}\quad \lambda \lambda_{S} \geq \lambda_{HS}^2  \ .
}
Now, requiring electroweak symmetry breaking through the CW mechanism means that we should 
set the couplings such that there is a flat direction in the potential. 
There are three options
\begin{enumerate} 
\item  if $\lambda_S(\mu_0) \approx 0$ then $S$ is a flat direction, with $\langle H \rangle = 0$, meaning that electroweak symmetry is unbroken. This case is, of course, ruled out.
\item  if $\sqrt{\lambda (\mu_0)\lambda_S(\mu_0)} = -\lambda_{HS}(\mu_0)$ there is a flat direction, which is a linear combination of $H$ and $S$, and leads to electroweak symmetry breaking with the Higgs-boson being a mixture of $H$ and $S$ quanta. This case was considered explicitly in Ref.~\cite{Sannino:2015wka}, and, to the perturbative order studied, is incompatible with experiments. 
\item if $\lambda(\mu_0) \approx 0$ then $H$ is a flat direction, with $\langle S \rangle = 0$. This is the case we pursue here, as in Ref.~\cite{Antipin:2013exa}.
\end{enumerate}

To compute the one-loop corrections to the Higgs-potential, we use the background field method, and take as background:
\ea{
H = \frac{1}{\sqrt{2}} \begin{pmatrix} 0 \\ \phi_h \end{pmatrix} \, , \quad S = 0 \ .
}
The one-loop corrections to the potential on any background and in the Landau gauge reads :
\be
\label{oneloopterm0}
V_1(\phi_h)&=&\frac{1}{2}\int \frac{d^4 k}{(2\pi)^4} \ Str \big[ \ln (k^2+M^2(\phi_h))\big]+c.t. 
\ee
where
$M^2(\phi_h)$ is the background dependent mass matrix and we defined the supertrace
\ea{
\label{Str}
Str\equiv \sum\limits_{\rm scalars}-\ 2\sum\limits_{\rm Weyl-fermions}+\ 3\sum\limits_{\rm vectors} \ .
}
In Ref.~\cite{Antipin:2013exa} we computed this quantity through cutoff-regularization. 
In this paper we will instead use dimensional regularization, to better connect with experimental data. As shown in Ref.~\cite{Einhorn:1992um}, the Veltman conditions for the cancellation of quadratic corrections 
to scalar self-energies follow from the quadratic part of the potential and are given by:
\be
\frac{1}{2}\frac{\partial^2 Str[M^2(\phi_i)]}{\partial \phi_i^2} \ \bigg|_{\mu_0}=0 \ ,
\label{Veltman}
\ee
where $\phi_i$ is any non-Goldstone background scalar field. We will investigate these relations in the end. For now, we may proceed as in the original work of Coleman and Weinberg by fine-tuning these divergence away.
Then after normalizing the vacuum energy to zero,
the one-loop corrections to the potential in Landau gauge and the $\overline{\rm MS}$-scheme read (see e.g.~\cite{Martin:2001vx}):
\ea{
V_1(\phi_h) = Str\left [ \frac{M^4(\phi_h)}{4} \left (\log\frac{M^2(\phi_h)}{\mu_0^2} - C_i \right ) \right ] \ ,
}
where $C_i$ are constants which for scalar and fermions read $C_i = 3/2$ and for vector bosons read $C_i = 5/6$. For this model, and for $\lambda(\mu_0) \approx 0$, the mass-matrix on the Higgs-background reads:
\be
\label{SMmassmatrix}
\frac{M^2(\phi_h)}{\phi_h^2} =\text{diag} \left \{ 
\frac{1}{4}g^2, \ \frac{1}{4}g^2, \ \frac{1}{4}(g^2+g^{\prime 2}), \ \frac{1}{2}y_t^2, \ \frac{1}{2}y_t^2 , \lambda_{HS}\right\} \equiv \mathcal{M}^2 \ ,
\ee
where the entries correspond respectively to 
the $W^+$, $W^-$ and $Z$ vector bosons, two top quark color multiplets in the Weyl basis, and the new singlet boson $S$.
At the last equality we have defined the field-independent matrix $\mathcal{M}^2$, which is possible since there are no explicit tree-level renormalized mass-terms in the theory considered. However, if the new explicit mass-terms for the fermions will appear this expression will be modified, but will not alter the Veltman conditions.
The one-loop potential simplifies to
\ea{
V_1(\phi_h) = A\, \phi_h^4 + B \,\phi_h^4 \, \log\left( \frac{\phi_h^2}{\mu_0^2} \right )\ ,
}
with
\ea{
A = \frac{1}{64\pi^2} Str \left [\mathcal{M}^4 \left (\ln \mathcal{M}^2 - C_i \right) \right ] \ , \quad 
B= \frac{1}{64\pi^2} Str \mathcal{M}^4 \ .
} 
Whenever $B > 0$, the potential has a non-zero vacuum expectation value at:
\be
\log\frac{\langle \phi_h \rangle^2}{\mu_0^2}=-\frac{1}{2}-\frac{A}{B} \ ,
\ee
and the Higgs-boson gains a renormalized mass at one-loop reading:
\be
\label{oneloopmass}
m^2_h(\mu_0)= 8 B \langle \phi_h \rangle^2 \ . 
\ee
The other scalar field $S$ obtains a tree-level mass through the coupling to the Higgs, reading:
\ea{
m_S^2(\mu_0) = \lambda_{HS}(\mu_0) \langle \phi_h \rangle^2 \ .
\label{Stree}
}
It is convenient to express the relation between $\langle \phi_h \rangle $ and $\mu_0$ as
\ea{
\mu_0= \langle \phi_h \rangle \,   e^{\frac{1}{4} + \frac{A}{2B}} = v  \,   e^{\frac{1}{4} + \frac{A}{2B}} 
\label{mu0}
}
where at the last equality we identified $\langle \phi_h \rangle $ with the experimentally measured electroweak vacuum expectation value, denoted $v$, with the fixed value:
\ea{ v = 246.21(9) \text{ GeV} \ . }
\change{
The value of $\mu_0$ may not be much larger than $v$ for the perturbative expansion in $\log(v/\mu_0)$ to be valid.}
The quantities $A$ and $B$ are easily computed. Defining first the quantity:
\ea{
\mathcal{G}^4 = 3g^4+2 g^2 g'^2+g'^4 \ ,
}
we find
\ea{
B =& \frac{1}{64 \pi^2} \left [ 
\frac{3}{16}\mathcal{G}^4-N_c y_t^4+ \lambda_{HS}^2 \right ] \ ,
\\
A =& \frac{1}{64 \pi^2} \left [ \frac{2}{16} \mathcal{G}^4 
+ \frac{6g^4}{16} \log \frac{g^2}{4}
+ \frac{3(g^2+g'^2)^2}{16} \log \frac{g^2+g'^2}{4}
- N_c y_t^4 \log \frac{y_t^2}{2} + \lambda_{HS} \log \lambda_{HS} \right ] 
- \frac{3}{2} B
\, ,
}
where $N_c=3$ is the number of colors.
In the $\overline{\rm MS}$-scheme the tree-level coupling constant values at the top-mass scale read:
\ea{
g(M_t) = 0.64(8) \, , \ g'(M_t) = 0.35(0) \, , \ y_t(M_t) = 0.99(5) \, ,
\label{MStree}
}
with $M_t = 173.34$ GeV. 
\change{There remains only one undetermined parameter, which is the portal coupling $\lambda_{HS}$.
To fix its value we must determine the Higgs-pole mass and identify it with the experimental value.
This will self-consistently provide the value of $\mu_0$ and predict the running mass of $S$.
In turn, we will identify the pole-mass of $S$ with $750$ GeV. Thus, in the next section we go through the pole-mass derivations.}

\section{Physical mass predictions}
\label{polemass}

The effective potential used to compute the scalar masses, by definition, is evaluated at zero external momentum while we need to calculate the physical (pole) mass of the scalars  which is computed at the momentum equal to the pole mass itself:
\begin{equation}
\label{dilpole1}
M_{pole}^2= m_0^{2}+\Pi (p^2=M_{pole}^2) \ .
\end{equation}
where we denoted the scalar self-energy function by $\Pi(p^2)$ and $m_0$ is the tree-level mass of the scalar.
Now, we can use the definition of the running mass 
\be
\label{runmass}
m_{\rm run}^2\equiv m_0^2+\Pi (p^2=0) \ ,
\ee
 to express the pole mass as:
\begin{equation}
M_{pole}^2= m_{\rm run}^2 +\Pi (p^2=M_{pole}^2)-\Pi (p^2=0)\equiv m_{\rm run}^2 +\Delta \Pi  \ .
\label{pole}
\end{equation}
and we observe a shift due to the scalar self-energy $\Delta \Pi$ when we convert from running to pole mass. In Eq.~\eqref{pole}, $m_{\rm run}^2$ and $\Delta \Pi $ both depend on the RG scale $\mu_0$ in such a way that the resulting pole mass $M_{pole}$ is RG independent \cite{Casas:1994us}. Namely,
\be
\frac{d \Delta \Pi}{d \log \mu_0}=- 2 \gamma M_{pole}^2 \qquad \Longrightarrow \qquad \Delta \Pi =  2 \gamma M_{pole}^2 \log \frac{M_{pole}}{\mu_0} +{\rm const}
\ee
where $\gamma$ is the scalar field anomalous dimension. 
From the SM top Yukawa contribution $y_t$ (with $ \gamma_H = 3 y_t^2/(4\pi)^2)$ and from the mixed quartic $\lambda_{HS}$ we have the self-energy correction for the mass of  $h$ \cite{Casas:1994us, Gonderinger:2009jp}:
\begin{equation}
\Delta \Pi_{h}\approx \frac{3 y_t(\mu_0)^2}{16\pi^2}
\left\{(4M_{t}^2- M_h^2)\left[2-Z\left(\frac{M_{t}^2}{M_{h}^2}\right)
\right]+M_{h}^2 \log
\frac{M_{t}^2}{\mu_0^2}
\right\} +\frac{\lambda_{HS}(\mu_0) M_{S}^2}{8\pi^2} \left[Z\left(\frac{M_{S}^2}{M_{h}^2}\right)-2\right]
\label{poleH}
\end{equation}
where $M_t$, $M_S$ and $M_h$ are the physical pole masses of the top quark, scalar and Higgs boson, $\mu_0$ is the Coleman-Weinberg RG scale, to be determined self-consistently, and the function $Z(x)$ is defined as:
\begin{eqnarray}
Z(x)&=&\left\{
\begin{array}{ll}
2 A \:{\rm tan}^{-1}(1/A), & \mbox{if $x>1/4$}\\
A \:\log\left[(1+A)/(1-A)\right],&\mbox{if $x<1/4$}
\end{array}\right.\nonumber\\
A&\equiv&|1-4x|^{1/2} \  ,
\end{eqnarray}
which takes the limiting value $Z(x=\infty)= 2$. Now, fixing $M_t = 173.34$ GeV, $M_h = 126$ GeV, and $M_S = 750$ GeV,
we can get $\lambda_{HS}(\mu_0)$ iteratively by initially setting $\mu_0 \approx M_t$,
and solving the equation $M_h^2 = m_h^2(\mu_0) + \Delta \Pi_h (\mu_0)$.
This leads to 
\ea{
\lambda_{HS}(M_t) = 4.85 \ .
}
From this and Eq.~\eqref{mu0} we get a new estimate for $\mu_0$, giving
$\mu_0 = 188 \text{ GeV}$.
Since this value is approximately the same as $M_t$, we can stop the iterative process already, meaning that we can simply take Coleman-Weinberg RG scale to be $\mu_0 \approx M_t$, since the logarithmic change of $M_t$ to $188$~GeV is very small.
Although the above value of $\lambda_{HS}$ is large, it is still perturbative. We will comment on its Landau pole at larger scales in the end.

Analogously taking into account the, yet unspecified, Yukawa coupling(s) contribution(s) to the anomalous dimensions of the $S$, $\gamma_S$, from the new fermion(s) (which we will call collectively $\chi$) and contributions from the mixed quartic $\lambda_{HS} $ we have \cite{Gonderinger:2009jp}:
\begin{equation}
\Delta \Pi_{S}\approx \gamma_S
\left\{(4M_{\chi}^2- M_S^2)\left[2-Z\left(\frac{M_{\chi}^2}{M_{S}^2}\right)
\right]+M_{S}^2 \log
\frac{M_{\chi}^2}{\mu_0^2}
\right\} +\frac{\lambda_{HS}(\mu_0) M_{S}^2}{4\pi^2} \int_0^1{d x \log  \frac{M_h^2(1-x)+M_S^2 x^2}{M_h^2 (1-x)+M_S^2 x}} \ .
\label{poleS}
\end{equation}
where $M_\chi$ is the pole mass of the fermion $\chi$. Again identifying $S$ with a resonance at $750$ GeV we arrive at the equation for $\gamma_S$ and $M_\chi$
\ea{
M_S^2 = m_S^2 + \Delta \Pi_S (\gamma_S, M_\chi) = (750~\text{ GeV})^2 \ , 
\label{Spolemass}
}
leading to a further relation between the physical mass of the new fermion and its contribution to the anomalous dimension of $S$. The running mass $m_S$ is given by the Coleman-Weinberg prediction, Eq.~\eqref{Stree}.
Let us note that it is possible to add an explicit tree-level mass for $S$, which would not change any of the earlier results. It would, however, modify the  equation above, thus modifying  the constraints for the new fermionic sector. 

In the next section we turn our attention to the new fermionic sector and its possible relation to the recent diphoton excess, which will be used as input to constrain one of the two remaining parameters of the model, $\gamma_S$ and $M_\chi$.

\section{ Completing the model: Di-photon excess}
\label{predictions}

The ATLAS collaboration has collected 3.2 fb$^{-1}$ of data and reported an excess in 
the distribution of events containing a pair of photons, at the di-photon invariant mass $M\approx \Excess$
with $3.9\sigma$ statistical significance \cite{Atlas}.
The ATLAS data consists of about 14 events which appear in two energy bins,
suggesting a best-fit width of about 45~GeV ($\Gamma/M \approx \GM$).
The result is partially supported by the CMS collaboration \cite{CMS:2015dxe} which collected integrated luminosity of 2.6 fb$^{-1}$ and has reported
a modest excess of about 10 $\gamma\gamma$ events peaked at 760 GeV.
The best-fit has a narrow width 
and a local statistical significance of $2.6\sigma$. The cross sections in a di-photon invariant mass interval corresponding to the best-fit width values
can be estimated as \cite{Franceschini:2015kwy}
\be\label{excess}
\sigma(pp\to\gamma\gamma) \approx 
\left\{\begin{array}{lll}
(6\pm3)\fb~&\hbox{CMS}&\sqrt{s}=13\ \rm {TeV},\\
(10\pm3)\fb~&\hbox{ATLAS}&\sqrt{s}=13 \  {\rm TeV} .
\end{array}\right.
\ee

This value can be easily obtained in a weakly-coupled model, however, the total width, generated by the one-loop decays to pair of photons $\gamma\gamma$ and gluons $gg$ alone, is too small to fit the ATLAS observation.
The typical expression for a tree-level decay width is $\Gamma/M\sim y^2/4\pi$;
so the total width $\Gamma/M\approx \GM$ 
can be explained via a tree-level decay process if the relevant coupling $y\sim \mathcal{O}(1)$.

Let us reflect on these experimental results in view of our model.
\begin{itemize}
\item Our model has one extra real scalar $S$ with \change{a coupling to the Higgs, $\lambda_{HS}(M_t) = 4.85$}, required to explain the experimental value of the pole mass of the Higgs. We propose to identify the $S$-boson with the newly observed excess in the di-photon invariant mass around \Excess . 

\item  \change{Our analysis in Ref.~\cite{Antipin:2013exa} suggests that the new scalar should be coupled to new fermions, which we have so far generally modeled by the Yukawa sector  $Y_{ij} S \psi_i \psi_j$ in Eq.~\eqref{lagr}.}
 To keep the Higgs mass prediction intact, the new fermions $\psi_i$ should couple dominantly to $S$. Once the fermionic sector is specified, we have to find the physical (pole) mass of $S$ using Eq.~\eqref{pole} and compare it with the experimental excess around \Excess.

\end{itemize}

Let us illustrate how we can address the points above via the introduction of a new fermionic sector. We explore the consequences of adding a QCD-colored vector-like quarks, $\chi^i$, in the representation $r_i$  and carrying hypercharge, $Y^i = Q_\chi^i$. 
The additional contribution to the potential of the theory reads
\be
V &=& (y_\chi^{ij} S+ m_{\chi}^{ij}) \bar{\chi}_L^i \chi_R^j  + (y_{\chi q}^{ab} S+m_{\chi q}^{ab}) \bar{\chi}_L^a q_R^b   +y_{H\chi}^{a b} \bar{Q}_L^a \chi_R^b H +h.c.
\label{possibleY}
\ee
Clearly, the couplings $y_{\chi q}^{ab}, y_{H\chi}^{a b}$ and invariant bare masses $m_{\chi q}^{ab}$ are only allowed for a $\chi^i$-quarks which are triplets of QCD and are of down-type ($Q_\chi^i$=-1/3) or up-type ($Q_\chi^i$=2/3) while the couplings $y_\chi^{ij}$ and invariant bare masses $m_{\chi}^{ij}$ may contain additional non SM-like species of vector-like fermions.
As we discussed above, as long as new fermions will couple only to $S$, the one-loop prediction for the Higgs mass will not be affected which leads us to consider the $y_{H\chi}^{ab}\approx 0$ region of this model. 
Also, we assume that, when allowed, $\chi^a$ couples predominantly to the 3rd family of SM quarks so that $Q_L=(t_L \ b_L)^T$ is the SM doublet and $q_R=b_R$ or $t_R$ depending on whether the $\chi^a$ quark is bottom-like or top-like.

To be predictive, we start with the model featuring non-SM like quarks so that we have  $y_{\chi q}^{ab}=m_{\chi q}^{ab}=0$ and comment on the more general situation at the end. Furthermore, we go to the basis where $y_{\chi}^{ij}$ is diagonal.

The Yukawa's $y_{\chi}^i$ generate the $S gg$ and $S\gamma \gamma$ vertices via a $\chi^i$-loop 
and the production cross section of $S$ through gluon fusion times the branching ratio into di-photon for a specie $i$ is then given by \cite{Franceschini:2015kwy,Harigaya:2015ezk}
\footnote{ In addition, we can make a loop of $\chi$-quarks with two photons attached, generating another contribution to $\sigma (pp\to 2\gamma)$ and this process is also available for SM-like quarks.}:
\be
\sigma_i (pp\to\gamma\gamma)\approx \sigma_i (pp\to S) \times Br_i (S\to \gamma\gamma)\approx 10 \ {\rm fb} \times d(r_i) ^2 \times |y_{\chi}^{i}|^2 \times \bigg(\frac{Q_{\chi}^i}{2/3}\bigg)^4 \times \bigg(\frac{330 \ {\rm GeV}}{M_{\chi^i}}\bigg)^2 \ ,
\label{exp}
\ee 
where $d(r_i)$ is the dimension of the color-representation of the quark specie $\chi^i$, and $Q_\chi^i$ is its hypercharge.
To reproduce the observed excess around $750$ GeV in the di-photon invariant spectrum, from Eq.~\eqref{excess}, the total cross-section of $\sigma(pp\to\gamma\gamma)\sim 10 \ {\rm fb} $ is required, explaining the normalization.

\begin{figure}[tb]
\begin{center}
\includegraphics[width=0.8\textwidth]{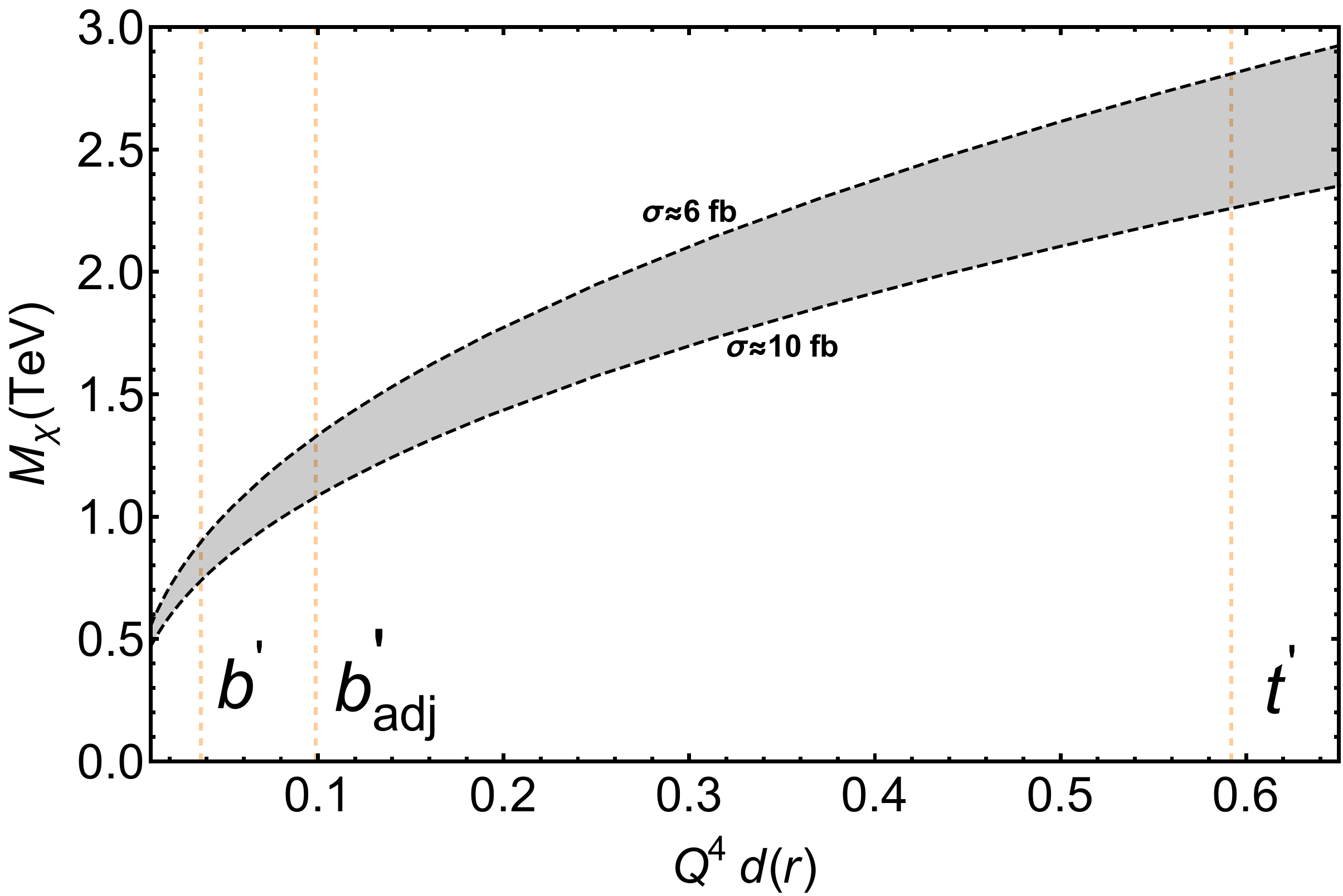}
\caption{ The mass of the new fermions $M_\chi$ as a function of electric charge $Q^4$ and dimension of the QCD representation $d(r)$. The orange gridlines correspond to the  SM-like $b^\prime$ (left), bottom-like quark in the adjoint representation of QCD $b^\prime_{\rm adj}$ (middle) and SM-like $t^\prime$ (right).}
\label{Br}
\end{center}
\end{figure}

In principle, for a given  $y_{\chi}^{i}$ and $m_\chi^i$ we need to evaluate the pole masses of the fermions $M_\chi^i$ and use Eq.~\eqref{poleS} to predict the pole mass $M_S$ of the scalar $S$. Since these numbers are unknown, we simply fix $M_S=750$ GeV assuming that there is a \emph{single} non SM-like quark $\chi$ with pole mass $M_{\chi}$ responsible for it.  
This new quark yields a one-loop contribution to the scalar anomalous dimension 
\ea{ 
\gamma_S = d(r) \frac{|y_{\chi}|^2}{(4\pi)^2} \ .
}
From Eq.~\eqref{exp} we have a relation between $y_\chi$ and $M_\chi$ which we use in Eq.~\eqref{Spolemass}. This allows us to relate $M_\chi$ to the electric charge and dimension of the QCD representation $d(r)$, which is shown in Fig.~\ref{Br}. \change{The shaded region corresponds to the combined data of ATLAS and CMS, i.e. $\sigma \approx 8 \pm 2 {\rm ~fb}$.
The orange vertical dashed-line in the middle is an example of a non-SM bottom-like quark, which is in the adjoint representation of QCD $b^\prime_{\rm adj}$ (predicting $M_\chi\approx {\rm \ 1.1-1.3 \ GeV}$ and $y_\chi\approx 1.64-1.56$). 

The results in Fig.~\ref{Br} are valid also for SM-like quarks as long as $y_\chi$ dominates over other Yukawa couplings. For example, the right orange vertical dashed-line corresponds to a SM-like $t^\prime$ predicting $M_\chi\approx 2.3-2.8 {\rm \ TeV}$, and $y_\chi\approx 2.3-2.2$, while the left orange vertical dashed-line corresponds to $b^\prime$ state with $M_\chi\approx 740-900$ GeV, and $y_\chi \approx 3.0-2.8$. In the following section we discuss the implications of these predictions.
}

\section{Predictions}
\label{predictions}

\change{
In the previous section we restricted the analysis, for simplicity, to the case where the Yukawa couplings in Eq.~\eqref{possibleY} other than $y_\chi$ are vanishing. This is the only possibility if the new quark does not carry SM-like charges, such as in the example of the adjoint $b^\prime_{\rm adj}$.
However, as also evident from the analysis, the new quark could carry SM-like charges, e.g. in  the $t^\prime$ and $b^\prime$ case. }
In that case (still restricting our analysis to $y_{H\chi} \approx 0$, disallowing $\chi\to q H$) we can have a decay of $\chi \to q S$ via the $y_{\chi q}$ Yukawa coupling, which predicts the following resonant contribution around $M_\chi$:
\be
pp\to \chi \chi \to qq SS \to qq 4\gamma \ (4\  \rm gluons) \ .
\ee
The coupling $y_{\chi q}$ will also contribute to the anomalous dimension of $S$ which, for the case of a single new quark reads ($d(r)=N_c$ for the fundamental QCD representation):
\be
\gamma_S = N_c \frac{ |y_{\chi}|^2 + 2 |y_{\chi q}|^2} {(4\pi)^2} \ .
\ee
 This leads to extra contributions in $\sigma(pp\to\gamma\gamma)$ and, generically, allows to fit the di-photon excess for smaller values of Yukawa couplings, making the model more weakly-coupled.

Another implication of the new heavy quarks is that since $S$-boson is a real scalar \cite{Ghosh:2015apa,Megias:2015ory,Falkowski:2015swt,Chakrabortty:2015hff,Chao:2015ttq,No:2015bsn,DiChiara:2015vdm,Cao:2015pto,Han:2015qqj,Dutta:2015wqh,Buttazzo:2015txu,McDermott:2015sck,Fichet:2015vvy}, below the mass of $M_\chi$ there can appear an effective gauge-invariant SM operator $S B_{\mu\nu}B^{\mu\nu}$ with $B_\mu$ the hypercharge gauge boson (in contrast to the pseudoscalar case, which instead couples to $\epsilon^{\mu\nu\rho\sigma}{B}_{\mu\nu}B_{\rho\sigma}$ \cite{Molinaro:2015cwg,Nakai:2015ptz,Low:2015qep,Matsuzaki:2015che, Bian:2015kjt,Jaeckel:2012yz,Volovik:2014nea,Volovik:2012qq}). This naturally implies  $S\to \gamma\gamma, ZZ, Z\gamma$ branchings that are roughly related as $1: 0.09 : 0.6$. In fact, we could also replace $S$ with a pseudoscalar without affecting much our main results. This model can in this sense be seen as the elementary (perturbative) counterpart of models of minimal composite electroweak dynamics investigated in \cite{Molinaro:2015cwg}. In the composite case one also predicts sizable decays into WW, which are not present here and thus distinguish the two.

\section{On perturbative naturalness and conclusions}
\label{conclusions}

\begin{figure}[bht]
\begin{center}
\includegraphics[width=0.49\textwidth]{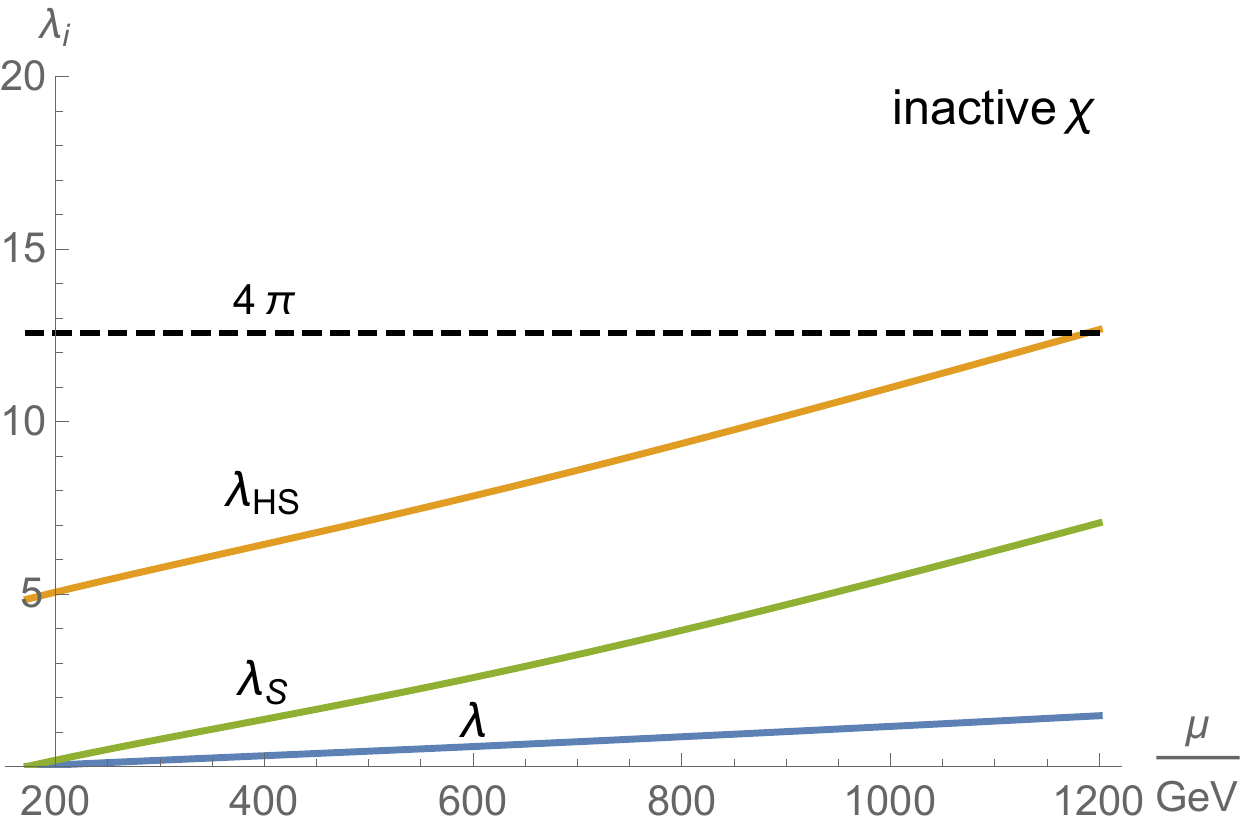}
\includegraphics[width=0.49\textwidth]{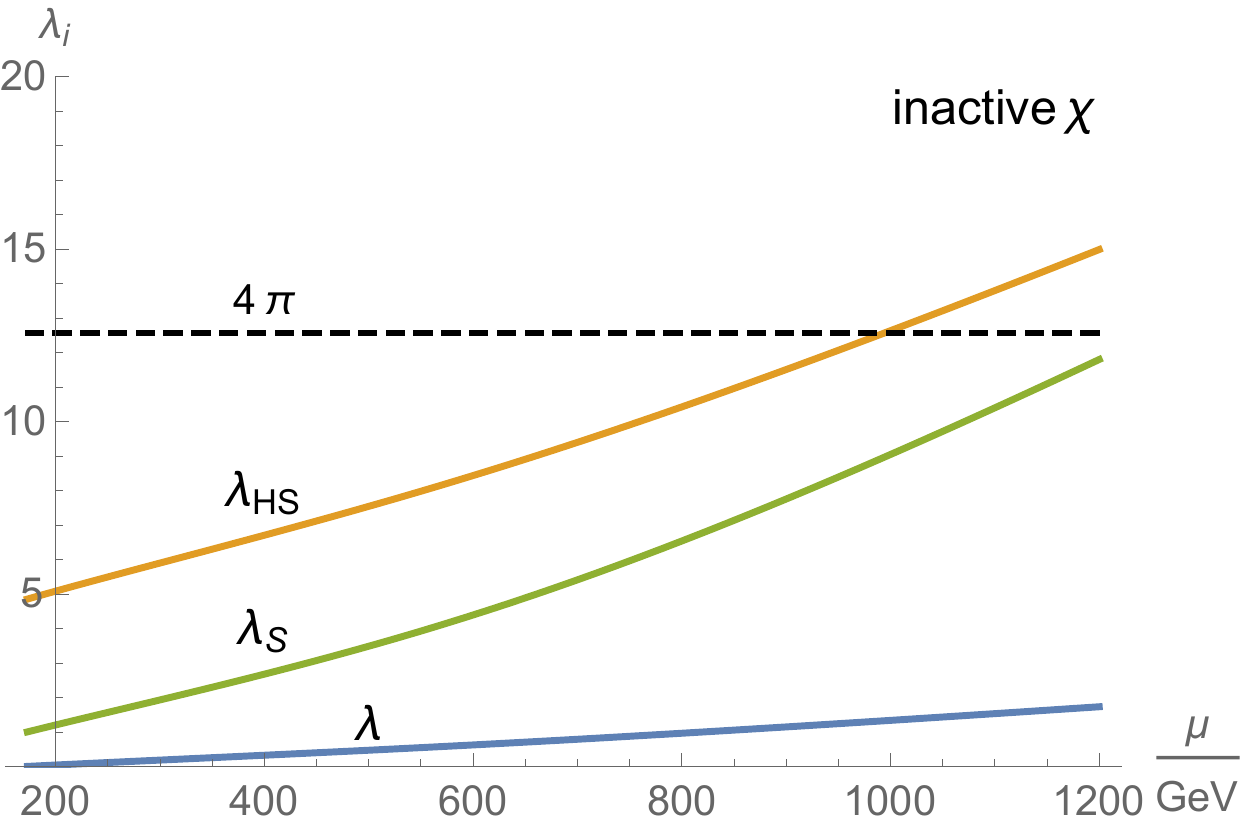}
\caption{Estimate of the running scalar couplings, assuming any dark coupling to the singlet scalar $S$ to be `inactive', in the sense of MS-renormalization. In the left panel, the initial value of $\lambda_S(M_t) = 0$, while in the right panel it is $\lambda_S(M_t) = 1$.}
\label{RG}
\end{center}
\end{figure}

One may worry about the perturbative validity of our predictions, since the portal coupling has a large value $\lambda_{HS}(M_t) = 4.85$. This value is still perturbative at the top-mass scale, since $\lambda_{HS}/4\pi < 1$. However, we must ensure that it does not run into non-perturbative values within the relevant energy-scales of the model. 
We consider therefore the RG equations of the scalar sector, given in the $\overline{\rm MS}$-scheme by:
\ea{
(4\pi)^2\frac{d\lambda}{d\ln\mu} =& \left(12 y_t^2-3 {g'}^2-9 g^2\right) \lambda-6 y_t^4+\frac{3}{8}\left[2g^4+ ({g'}^2+g^2)^2\right]+24 \lambda^2+ 2 \lambda _{{HS}}^2 \ , 
\\
(4\pi)^2\frac{d\lambda _{{HS}}}{d\ln\mu} =& \frac{1}{2}\left(12 y_t^2-3 {g'}^2-9 g^2\right) \lambda _{{HS}}+
3 \lambda _{{HS}}\left(4 \lambda+2 \lambda _{{S}}\right)+8 \lambda _{{HS}}^2  + \text{[Y]}_{\lambda_{HS}}\ ,    \\
(4\pi)^2\frac{d\lambda _{{S}}}{d\ln\mu} =& 8 \lambda _{{HS}}^2+18 \lambda _{{S}}^2 + \text{[Y]}_{\lambda_{S}} \ ,  
}
where the extra terms $\text{[Y]}_i$ are contributions from the new Yukawa coupling $y_\chi$, which is in this mass-independent schemes not `active' before the scale $\mu = m_\chi$, as well as other possible couplings to massive sterile/dark matter. We assume that none of these contributions are active at scales below 1 TeV.
Furthermore the running of the other SM couplings are unaffected by the new parameters.
With the assumption on the dark sector, we get an estimate of the running coupling as a function of RG scale, as shown in Fig.~\ref{RG}. 

From this analysis it is seen that the coupling $\lambda_S$, which was so far undetermined, cannot take values much bigger than one, since it otherwise drives the portal coupling $\lambda_{HS}$ to non-perturbative values at scales below 1 TeV.
The analysis also shows that the model becomes non-perturbative around 1-1.5 TeV.
This implies that the mass predictions for new quarks with charges giving large masses beyond 1.5 TeV cannot be trusted. We are thus lead to the conclusion that only the predictions for the low-mass new quarks are valid, such as the $b^\prime$ and the $b^\prime_{\rm adj}$. Furthermore it implies that beyond the new quark mass-scale, the model must either be extended or replaced by a strongly coupled theory. 
In the case of $b^\prime$ there is also the possibility to relax the condition $y_{H_\chi} \approx 0$. 
It could be interesting to consider this effect in the future.

Let us now discuss  the pertubative sensitivity of the model to quadratic divergencies both for the Higgs and the new scalar boson masses.  Having argued that the cutoff of the model is around 1-1.5 TeV one would naively expect to fine-tune the mass of the Higgs against these contributions and to a lesser extent also for $S$.  It turns out \cite{Antipin:2013exa} that, at the one-loop level,  for this model such a fine-tuning is alleviated because the Veltman conditions Eq.~\eqref{Veltman} are nearly satisfied. More precisely
\be
\label{VeltmanSinglet}
\frac{1}{2}\frac{\partial^2 Str[M^2(h)]}{\partial h^2} \ \bigg|_{\mu_0}= 
\frac{9}{4}g^2(\mu_0)+\frac{3}{4}g^{\prime 2}(\mu_0)-6y_t^2(\mu_0)+ \lambda_{HS}(\mu_0)=0.05
\ee
The mass of the Higgs in this model is thus \emph{perturbatively natural}~\cite{Antipin:2013exa}. The Veltman conditions for the new scalar mass are more model-dependent, and in any case its mass is already closer to the cutoff. 

To conclude, we have shown that extensions of the SM in which the Higgs-sector is replaced by the Coleman-Weinberg mechanism  can lead to the correct mass of the Higgs boson with minimal fine-tuning. We furthermore showed that this new scalar can explain the diphoton excess, through an associated new fermionic sector charged under $SU(3)_c$ and $U(1)_Y$. Finally we have suggested related signatures, which are predictive consequences of the model, and furthermore showed that the model has a natural cutoff at scales around 1-1.5 TeV. Thus we expect the LHC experiments soon to be able to test this alternative scenario for radiative breaking of the electroweak theory.

\acknowledgements
This work was partially supported by the Danish National Research Foundation DNRF:90 grant.

\end{document}